\documentclass[a4paper,11pt]{article}
\usepackage{pos}
\usepackage{amsmath}

\newcommand{\beqn}{\begin{eqnarray}}
\newcommand{\eeqn}{\end{eqnarray}}
\newcommand{\be}{\begin{equation}}
\newcommand{\ee}{\end{equation}}

\newcommand{\ba}{\begin{array}{c}}
\newcommand{\bat}{\begin{array}{cc}}
\newcommand{\ea}{\end{array}}

\newcommand{\bi}{\begin{itemize}}
\newcommand{\ei}{\end{itemize}}

\newcommand{\ket}{\,\rangle}
\newcommand{\bra}{\langle \,}

\newcommand{\Frac}[2]{\frac{\displaystyle #1}{\displaystyle #2}}

\newcommand{\cO}{{\cal O}}

\newcommand{\mF}{\mathcal{F}}

\newcommand{\mO}{\mathcal{O}}

\newcommand{\mT}{\mathcal{T}}

\newcommand{\gsim}{\stackrel{>}{_\sim}}

\newcommand{\comment}[1]{}

\title{Constraining resonances by using the electroweak effective theory}

\ShortTitle{Constraining resonances by using the EW effective theory}

\author*[a,\dag]{Ignasi Rosell}
\notes{\note{We wish to thank the organizers for the pleasant conference. This work has been supported in part by the Spanish Government and ERDF funds from the European Commission (FPA2016-75654-C2-1-P, FPA2017-84445-P, PID2019-108655GB-I00); by the Generalitat Valenciana (PROMETEO/2017/053); by the Universidad Cardenal Herrera-CEU (INDI18/11 and INDI19/15); by the EU STRONG-2020 project under the program H2020-INFRAIA-2018-1 [grant agreement no. 824093]; and by the STSM Grant from COST Action CA16108.
}}
\author[b]{Antonio Pich}
\author[c]{Juan Jos\'e Sanz-Cillero}

\affiliation[a]{Departamento de Matem\'aticas, F\'\i sica y Ciencias Tecnol\' ogicas, Universidad Cardenal Herrera-CEU, CEU Universities, 46115 Alfara del Patriarca, Val\`encia, Spain}
\affiliation[b]{IFIC, Universitat de Val\`encia -- CSIC, Apt. Correus 22085, 46071 Val\`encia, Spain}
\affiliation[c]{Departamento de F\'\i sica Te\'orica,  Universidad Complutense de Madrid, E-28040 Madrid, Spain}

\emailAdd{rosell@uchceu.es}
\emailAdd{pich@ific.uv.es}
\emailAdd{jjsanzcillero@ucm.es}

\abstract{In the light of the mass gap between Standard Model (SM) states and possible new particles, effective field theories are a suitable approach. We take on the non-linear realization of the electroweak symmetry breaking: the electroweak effective theory (EWET), also known as Higgs effective field theory (HEFT) or electroweak chiral Lagrangian (EWChL). At higher scales we consider a resonance electroweak Lagrangian, coupling SM fields to resonances. Integrating out these resonances and assuming a well-behaved high-energy behavior, some of the bosonic low-energy constants are determined or constrained in terms of resonance masses. Present experimental bounds on these low-energy constants allow us to push the resonance mass scale to the TeV range, $M_R \gsim 2\,$TeV, in good agreement with previous estimations.}

\FullConference{%
  40th International Conference on High Energy physics - ICHEP2020\\
  July 28 - August 6, 2020\\
  Prague, Czech Republic (virtual meeting)
}

\begin{document}

\maketitle

\section{Introduction}

The discovery of the Higgs and the non-observation of new particles has confirmed the Standard Model (SM) as our paradigm. Consequently, there is a mass gap between the SM and possible new physics (NP) fields. This gap supports a bottom-up approach, {\it i.e.}, the use of effective field theories to analyze systematically the low-energy data to search for fingerprints of heavy scales.

In this approach the low-energy constants (LECs), or Wilson coefficients, are free parameters from the effective point of view. Although they encode the information about the heavy scales, the structure
of the effective Lagrangian depends on the light-particle content (SM states in this case), the symmetries and the power counting. The power counting depends on the chosen scheme to introduce the Higgs field~\cite{Buchalla:2016bse}. The more common linear realization of the electroweak symmetry breaking (EWSB) is a first possibility: the SM effective field theory (SMEFT), where one assumes the Higgs to be part of a doublet together with the three electroweak (EW) Goldstones, as in the SM, and the Lagrangian is organized as an expansion in canonical dimensions, being its leading-order (LO) approximation the dimension-four SM Lagrangian. However, we adopt here the more general non-linear realization: the EW effective theory (EWET), also known as Higgs effective field theory (HEFT) or EW chiral Lagrangian (EWChL). In this second option one does not assume any specific relation between the Higgs and the EW Goldstones and an expansion in generalized momenta is followed. The LO Lagrangian is given in this case by the 
purely fermionic and gauge boson parts of the SM Lagrangian plus the $\cO(p^2)$ operators introducing the Higgs and the EW Goldstone interactions. It is interesting to stress that the SMEFT is a particular case of the more general EWET framework.

At shorter distances, heavy-mass resonances are introduced by using a phenomenological Lagrangian which follows the non-linear realization of the EW symmetry, {\it i.e.}, respecting the symmetries and observing the chiral expansion  of the EWET. 

As a result, two effective Lagrangians are considered here: the EWET at low energies, with only the SM fields, and the EW resonance theory at high energies, with the SM particles plus heavy resonances. By integrating out the resonances we can match both Lagrangians; that is, one can determine the EWET LECs in terms of resonance parameters. Assuming a good short-distance behavior is an important ingredient in this process. First of all, it is required in order to be able to consider the resonance theory as a good interpolation between the low- and the high-energy regimes. Secondly, these constraints are very convenient to reduce the number of resonance parameters and they allow us to get determinations or bounds in terms of only resonance masses. 

The ultimate aim is to combine current experimental bounds on the bosonic EWET LECs with their determinations or limits in terms of resonance masses, in order to 
constrain the NP scales~\cite{PRD,Pich:2015kwa}. 

\section{The effective Lagrangians}

The EWET Lagrangian is organized as an expansion in generalized momenta~\cite{Weinberg,Buchalla,lagrangian}:
\begin{eqnarray}
\mathcal{L}_{\mathrm{EWET}} &=& \sum_{\hat d\ge 2}\, \mathcal{L}_{\mathrm{EWET}}^{(\hat d)}\,. \label{EWET-Lagrangian0}
\end{eqnarray}
As it has been stressed previously, operators are not ordered following their canonical dimensions, one uses instead the chiral dimension $\hat d$, which indicates their infrared behavior at low momenta~\cite{Weinberg}. As a consequence, in this expansion loops are renormalized order by order. In Refs.~\cite{PRD,lagrangian} one can find the building blocks used to construct operators invariant under the electroweak symmetry group and the power-counting rules determining their chiral dimensions. For this work, the relevant bosonic part of the LO EWET Lagrangian is given by\footnote{An alternative notation $a=\kappa_W$, $b=c_{2V}$ is used also in the literature.}
\begin{eqnarray}
\Delta \mathcal{L}_{\mathrm{EWET}}^{(2)} &=&  \frac{v^2}{4}\,\left( 1 +\Frac{2\,\kappa_W}{v} h + \frac{c_{2V}}{v^2} \,h^2\right) \bra u_\mu u^\mu\ket_2    \, , \label{LOEWET_lagrangian}
\end{eqnarray}
being $h$ the Higgs field, $u_\mu$ the tensor containing one covariant derivative of the  the EW Goldstones and $\langle\cdots\rangle_2$ indicating an $SU(2)$ trace. Note that $\kappa_W$ parametrizes the
$hWW$ coupling.
The relevant $P$-even bosonic NLO EWET Lagrangian from a current phenomenological point of view reads~\cite{lagrangian}:\footnote{ 
These $\mF_j$ are related to the $a_i$ couplings of the Higgsless Longhitano Lagrangian~\cite{Longhitano} in the form $a_i=\mF_i$ for $i=1,4,5$, $a_2-a_3=\mF_3$.}
\begin{eqnarray}
\Delta \mathcal{L}_{\mathrm{EWET}}^{(4)} & =&  \sum_i  \mF_i\, \mO_i \,=\,
 \Frac{\mF_1}{4}\,\bra {f}_+^{\mu\nu} {f}_{+\, \mu\nu}- {f}_-^{\mu\nu} {f}_{-\, \mu\nu}\ket_2+
 \Frac{i\,\mF_3 }{2} \,\bra {f}_+^{\mu\nu} [u_\mu, u_\nu] \ket_2 \nonumber \\ &&\qquad \qquad
+ \mF_4 \bra u_\mu u_\nu\ket_2 \, \bra u^\mu u^\nu\ket_2 +
\mF_5 \bra u_\mu u^\mu\ket_2 \, \bra u_\nu u^\nu\ket_2   \, .\phantom{\frac{1}{2}} \label{EWET_lagrangian}
\end{eqnarray}
In the SM, $\kappa_W=c_{2V}=1$ and $\mathcal{F}_{1,3,4,5}=0$. The $W^\pm$ and $Z$ self-energies are sensitive to the operator $\mathcal{O}_1$ and, therefore, can be accessed through the measurement of the oblique parameter $S$~\cite{Peskin_Takeuchi}. Operators $\mathcal{O}_{1,3}$ and $\mathcal{O}_{1,3-5}$ contribute, respectively, to the trilinear and quartic gauge couplings.

The EWET power counting is not directly applicable to the resonance theory. However, the Lagrangian can be constructed in a consistent phenomenological way, {\it \`a la} Weinberg~\cite{Weinberg}, where the resonance electroweak theory interpolates between the low- and the high-energy regimes: one generates the appropriate low-energy predictions and a given high-energy behavior is imposed. Bearing in mind that we are interested in the resonance contributions to the bosonic $\cO(p^4)$ EWET LECs, only $\cO(p^2)$ operators with up to one bosonic resonance $R$ are required at tree-level~\cite{lagrangian}, standing $R$ for any of the four possible $J^{PC}$ bosonic states with quantum numbers $0^{++}$ (S), $0^{-+}$ (P), $1^{--}$ (V) and $1^{++}$ (A). The relevant resonance Lagrangian can be found in Refs.~\cite{PRD,lagrangian}.

If the heavy resonances are integrated out from the resonance Lagrangian, one recovers the EWET Lagrangian (\ref{EWET_lagrangian}) with explicit values of the LECs in terms of resonance parameters. As it has been explained previously, short-distance constraints are fundamental here. We note that without them one would determine the four EWET LECs of (\ref{EWET_lagrangian}) in terms of seven resonance couplings and the resonance masses. The following high-energy constraints have been assumed~\cite{PRD,lagrangian}:
\begin{enumerate}
\item Well-behaved form factors. The two-Goldstone and Higgs-Goldstone matrix elements of the axial and vector currents can be studied through the 
vector and axial-vector form factors. Assuming that these four form factors vanish at $s\to \infty$, we can find four constraints.
\item Weinberg Sum Rules (WSRs). The $W^3B$ correlator is an order parameter of the EWSB. In asymptotically-free gauge theories it vanishes at short distances as $1/s^3$~\cite{Bernard:1975cd}, implying two superconvergent sum rules~\cite{WSR}: the 1st WSR (vanishing of the $1/s$ term) and the 2nd WSR (vanishing of the $1/s^2$ term). While the 1st WSR is supposed to be also fulfilled in gauge theories with nontrivial ultraviolet (UV) fixed points, the validity of the 2nd WSR depends on the particular type of UV theory considered~\cite{1stWSR}.
\end{enumerate}

\section{Phenomenology}

\begin{table}[tb]  
\begin{center}
\renewcommand{\arraystretch}{1.2}
\begin{tabular}{|r@{$\,<\,$}c@{$\,<\,$}l|c|c||r@{$\,<\,$}c@{$\,<\,$}l|c|c| }
\hline
\multicolumn{3}{|c|}{LEC} & Ref. & Data & \multicolumn{3}{|c|}{LEC} & Ref. & Data \\ 
\hline \hline
$0.89$ & $\kappa_W$ & $1.13$  & \cite{deBlas:2018tjm}
& LHC  &
$-0.06$&$\mathcal{F}_3$ &$0.20$&\cite{Almeida:2018cld} & LEP \& LHC  \\ \hline 
$-1.02$ & $c_{2V}$ & $2.71$ & \cite{ATLAS:2019dgh} & LHC &
$-0.0006$&$\mathcal{F}_4$&$0.0006$&\cite{Sirunyan:2019der} & LHC  \\ \hline
 $-0.004$  &$ \mathcal{F}_1$& $0.004$  & 
 \cite{Tanabashi:2018oca} & LEP & 
$-0.0010$&$\mathcal{F}_4+\mathcal{F}_5$&$0.0010$ &  \cite{Sirunyan:2019der} & LHC  \\  \hline
\end{tabular}
\caption{{\small
Current experimental constraints on bosonic EWET LECs, at 95\% CL~\cite{PRD}.}} \label{exp}
\end{center}
\end{table}

\begin{figure*}[!t]
\begin{center}
\begin{minipage}[c]{7.0cm}
\includegraphics[width=7.0cm]{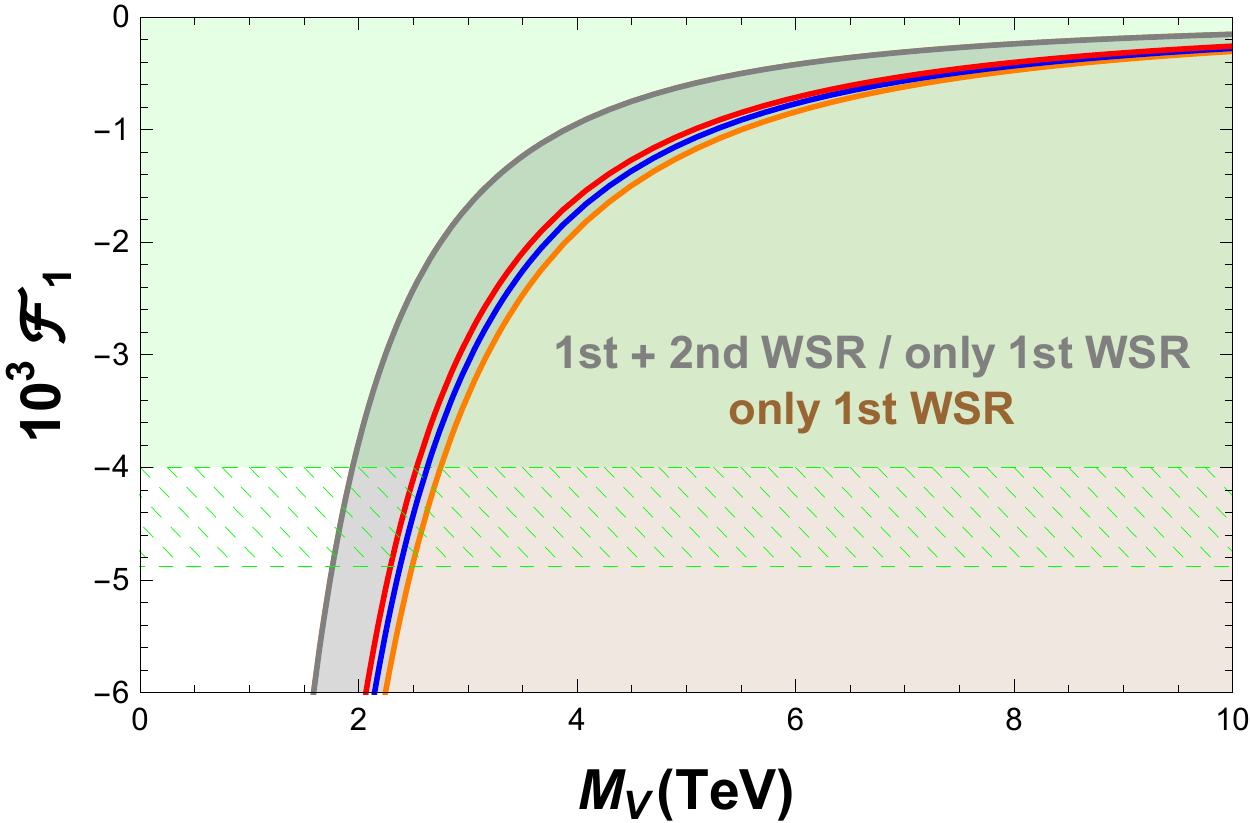}  
\end{minipage}
\hskip .5cm
\begin{minipage}[c]{7.0cm}
\includegraphics[width=7.0cm]{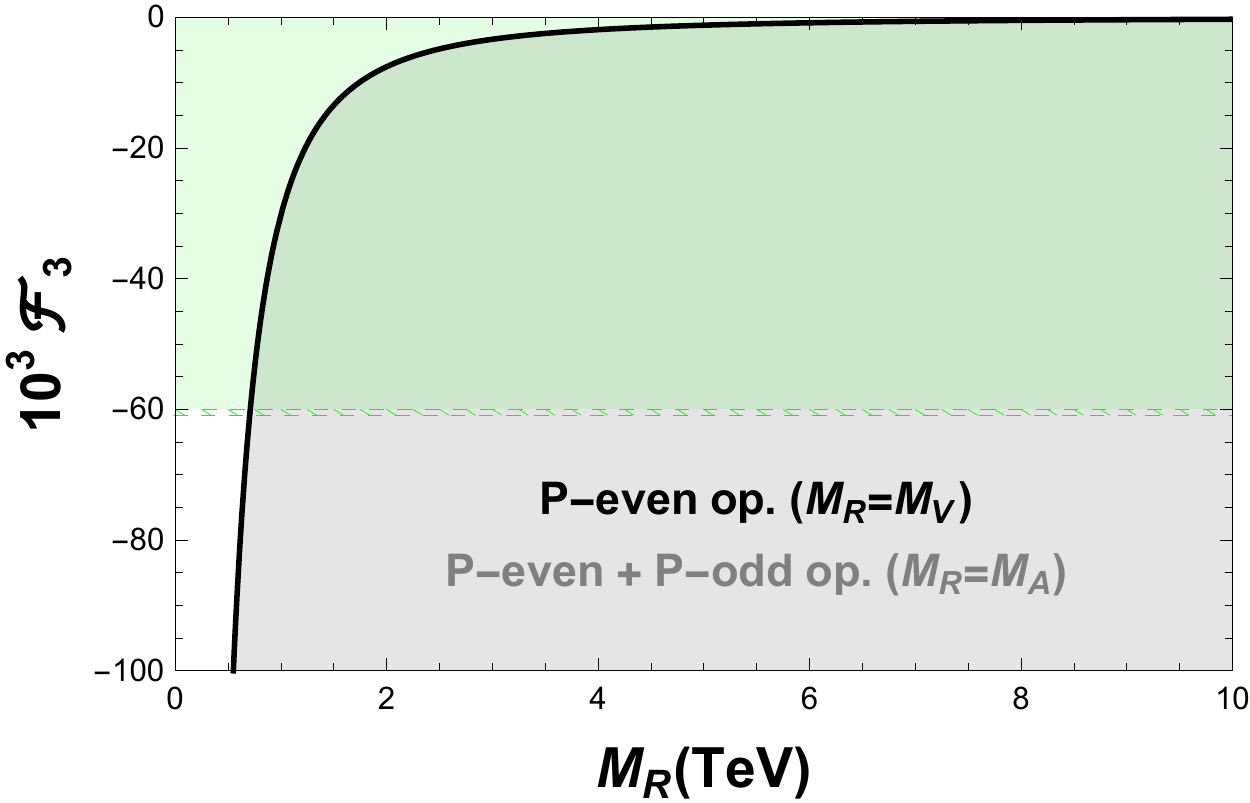} 
\end{minipage}
\\[8pt]
\begin{minipage}[c]{7.0cm}
\includegraphics[width=7.0cm]{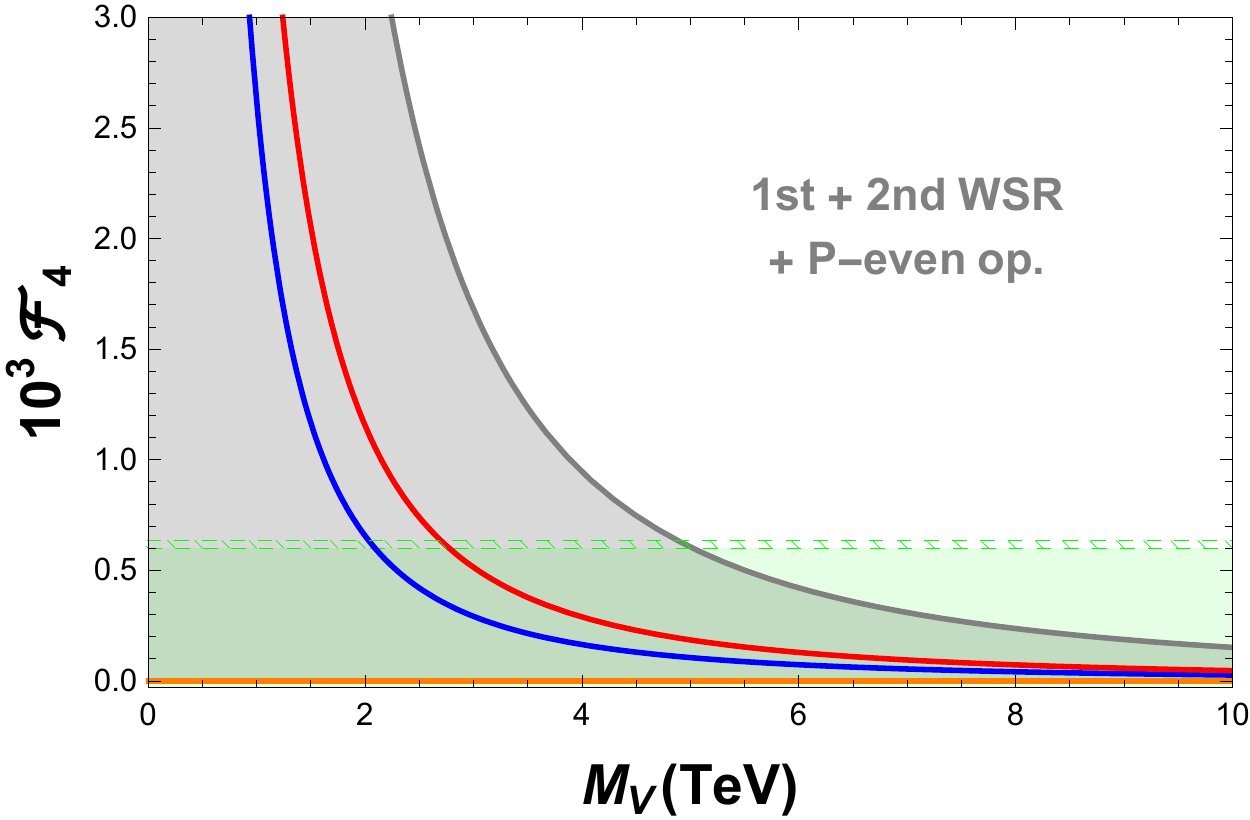} 
\end{minipage}
\hskip .2cm
\begin{minipage}[c]{7.0cm}
\includegraphics[width=7.0cm]{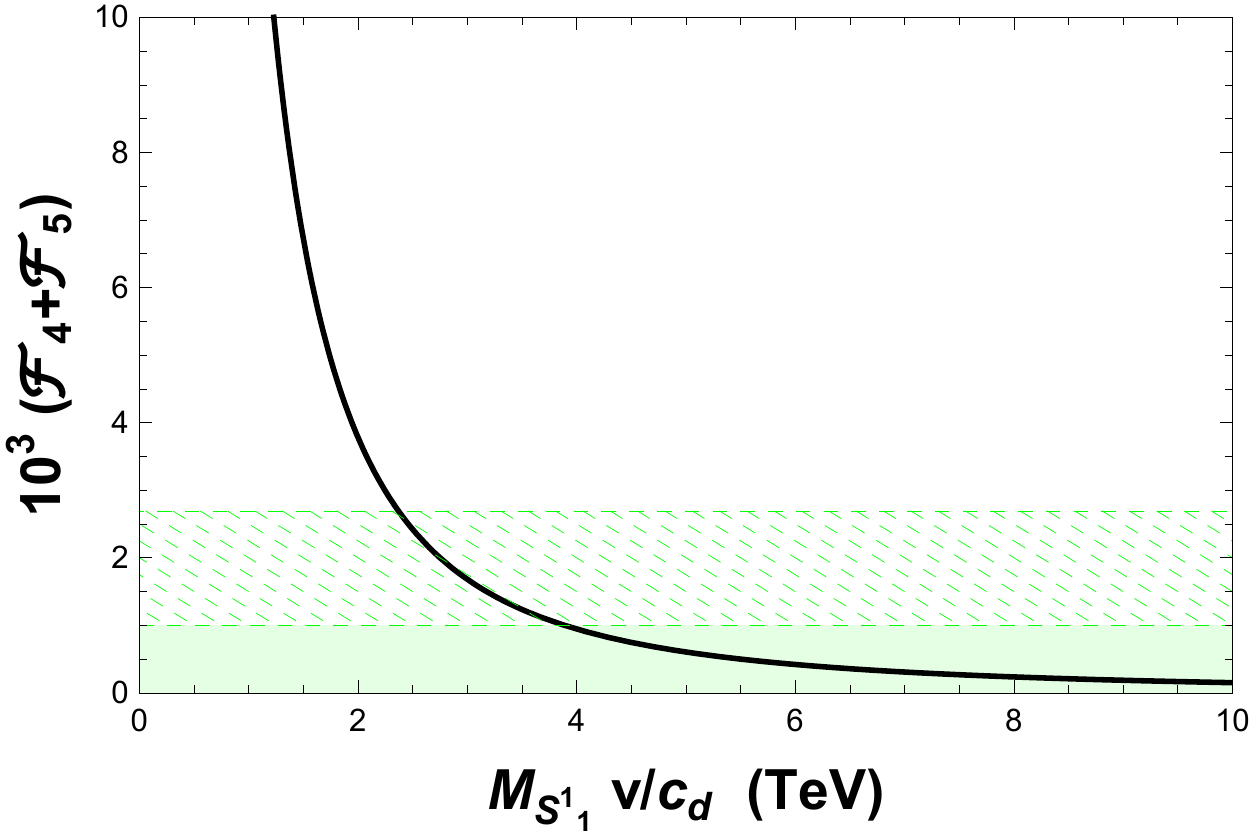}
\end{minipage}
\end{center}
\caption{{\small Prediction of the EWET LECs $\mF_1$, $\mF_3$, $\mF_4$ and $\mF_4+\mF_5$ of (\ref{EWET_lagrangian}) as functions of the corresponding resonance mass ($M_V$, $M_A$ or $M_{S_1^1} \,v/c_d$).
The 95\%-CL experimentally allowed regions given in Table~\ref{exp} are covered by the green areas, and they are further extended by dashed green bands accounting for our estimation of the one-loop running uncertainties. If the prediction of the LECs depend on both $M_V$ and $M_A$, the gray and/or brown regions cover all possible values for $M_A>M_V$. If the 2nd WSR has been assumed, we indicate it explicitly in the plot, with the corresponding lines for the limit $M_A=M_V$ (orange), $M_A=1.1\,M_V$ (blue), $M_A=1.2\,M_V$ (red) and $M_A\to \infty$ (dark gray). In the case without the 2nd WSR, the prediction for $\mF_1$ is given by the gray and brown regions. In case of using only the even-parity operators, it is indicated.
}} 
\label{plots1}
\end{figure*}

The use of the high-energy constraints explained at the end of the former section has allowed us to determine or bound the LECs $\mathcal{F}_{1,3,4,5}$ in terms of resonance masses~\cite{PRD}. On the other hand, the strongest experimental constraints on $\mathcal{F}_{1,3,4,5}$ are shown in Table~\ref{exp}~\cite{PRD}. We put together all this information in Figure~\ref{plots1}: the predictions of these LECs as functions of the relevant heavy resonance masses together with the regions allowed by the experimental constraints (green areas in the plots).

We display the dependence of $\mF_1$ on $M_V$ in the top-left plot in Figure~\ref{plots1}. If the 1st WSR is assumed, the dark gray curve shows the predicted upper bound $\mF_1 < -v^2/(4 M_V^2)$. Therefore, and if only the 1st WSR is obeyed, the whole region below this line (gray and brown areas) would be theoretically allowed. If one accepts moreover the validity of the 2nd WSR, $\mF_1$ is determined as a function of $M_V$ and $M_A>M_V$, being the dark gray curve the limit $M_A\to\infty$. The values of $\mF_1$ for some representative axial-vector masses are shown in the red ($M_A=1.2\, M_V$), blue ($M_A=1.1\, M_V$) and orange (the limit $M_A=M_V$) curves; being the orange line the lower bound $\mF_1=-v^2/(2M_V^2)$. This range of $M_A\sim M_V$ corresponds actually to the most plausible scenario~\cite{ST}. Then, and if both the 1st and the 2nd WSR are followed, only the gray region  would be theoretically allowed. 

For $\mF_3$ the WSRs do not play any role. Considering only $P$-even operators, one gets $\mF_3 = -v^2/(2 M_V^2)$ and we show this theoretical prediction by the black curve in the top-right plot of Figure~\ref{plots1}. If we add possible $P$-odd contributions, only an upper bound $\mF_3 < -v^2/(2 M_A^2)$ is found, which is represented by the same curve but this time with $M_R=M_A$. Therefore, the whole region below this line (gray area) would be allowed in the most general case.

If we assume the two WSRs and consider only $P$-even operators, $\mF_4$ is determined in terms of $M_V$ and $M_A$. We show in the bottom-left panel in Figure~\ref{plots1} the theoretical allowed values for this LEC as function of $M_V$. The upper bound (dark gray curve) is obtained at $M_A\to\infty$. Thus, the theoretically allowed region is the gray area below that curve. The values of $\mF_4$ for some representative axial-vector masses are shown in the red ($M_A=1.2\, M_V$), blue ($M_A=1.1\, M_V$) and orange (the limit $M_A=M_V$) curves again. Notice that the vector and axial-vector contributions have different signs and exactly cancel each other in the equal-mass limit.

Independently of any assumptions related to WSRs or $P$-odd operators, the contributions from vector and axial-vector resonance exchanges cancel exactly in the combination $\mF_4 + \mF_5 = c_d^2/(4 M_{S^1_1}^2)$ (being $c_d$ the $S_1WW$ coupling). This clean prediction of $\mF_4 + \mF_5$ is shown by the black curve in the bottom-right plot of Figure~\ref{plots1}, as function of $M_{S^1_1} v/c_d$.


Our tree-level predictions from resonance exchange are expected to apply at a scale around the resonance masses, whereas the experimental constraints on the LECs in Table~\ref{exp} have been obtained at lower scales. Taking into account these different scales and the known one-loop running of these LECs~\cite{Guo:2015isa}, the running contributions can be estimated and are indicated in Figure~\ref{plots1} with the dashed green bands that enlarge the experimentally allowed regions. These are of order $1/(4\pi^2) \!\sim\! 10^{-3}$ and depend on $\kappa_W$ and $c_{2V}$ of (\ref{LOEWET_lagrangian}), whose experimental constraints are also given in Table~\ref{exp}. 

The main conclusion is that Figure~\ref{plots1} pushes the resonance mass scale to the TeV range, $M_R \gsim 2\,$TeV, in good agreement with our previous theoretical estimates of Ref.~\cite{ST}, based on a NLO calculation of the $S$ and $T$ oblique parameters. The principal results are the following ones:
\begin{itemize}
\item The oblique $S$-parameter gives the most precise LEC determination ($\mF_1$), implying the lower bounds $M_{V,A}\gsim 2$~TeV (95\% CL).
\item The triple gauge couplings provide a weaker limit on $\mF_3$, which translates in the softer constraint $M_{V,A}\gsim 0.5$~TeV (95\% CL).  
\item For BSM extensions with only P-even operators and obeying both WSRs, the  bounds on $\mF_4$ constrain the mass of the vector resonance to $M_V\gsim 2$~TeV if $M_A/M_V>1.1$ (95\% CL).
\item The limit on $\mF_4+\mF_5$ implies that the singlet scalar resonance would have a mass $M_{S_1^1}\gsim 2$~TeV (95\% CL), assuming a  $S_1^1 WW$ coupling close to the $hWW$ one ($c_d\sim v$). 
\end{itemize}
Experimental constraints start already to be competitive. Much more precise information will be eventually got using this kind of analysis, once new data from the upgraded LHC runs be available.

\end{document}